\documentclass[preprint]{aastex}
\begin{document}

\title{The origin of complex organic molecules in prestellar cores}


\author{C. Vastel\altaffilmark{1,2}}
\affil{Universit\'e de Toulouse, UPS-OMP, IRAP, Toulouse, France}
\affil{CNRS, IRAP, 9 Av. colonel Roche, BP 44346, 31028 Toulouse Cedex 4, France}
\email{cvastel@irap.omp.eu}
\author{C. Ceccarelli\altaffilmark{3,4} and B. Lefloch\altaffilmark{3,4}}
\affil{Univ. Grenoble Alpes, IPAG, F-38000 Grenoble, France} 
\affil{CNRS, IPAG, F-38000 Grenoble, France} 
\author{R. Bachiller\altaffilmark{4}}
\affil{Observatorio Astron\'omico Nacional (OAN, IGN). Calle Alfonso
  XII,3. 28014 Madrid, Spain} 
    \date{Received - ; accepted -}

\begin{abstract}
Complex organic molecules (COMs) have been detected in a variety of
  environments, including cold prestellar cores. Given the low
  temperature of these objects, these last detections challenge
  existing models. We report here new observations towards the
  prestellar core L1544. They are based on an unbiased spectral survey
  of the 3mm band at the IRAM-30m telescope, as part of the Large
  Program ASAI. The observations allow us to provide the full census
  of the oxygen bearing COMs in this source. We detected tricarbon
  monoxide, methanol, acetaldehyde, formic acid, ketene, and propyne
  with abundances varying from $5\times10^{-11}$ to
  $6\times10^{-9}$. The non-LTE analysis of the methanol lines shows
  that they are likely emitted at the border of the core, at a radius
  of $\sim 8000$ AU where T $\sim$ 10 K and n$_{H_2}$ $\sim 2\times$
  10$^{4}$ cm$^{-3}$. Previous works have shown that water
  vapour is enhanced in the same region because of the photodesorption
  of water ices. We propose that a non-thermal desorption
    mechanism is also responsible for the observed emission of
    methanol and COMs from the same layer. The desorbed oxygen
  and a tiny amount of desorbed methanol and ethene are enough
  to reproduce the abundances of tricarbon monoxide, methanol,
  acetaldehyde and ketene measured in L1544. These new findings open
  the possibility that COMs in prestellar cores originate in a similar
  outer layer rather than in the dense inner cores, as previously
  assumed, and that their formation is driven by the non-thermally desorbed species.\end{abstract}



\section{Introduction}

Interstellar complex organic molecules (hereinafter COMs) are defined in
the astrochemical literature as organic molecules containing at least
six heavy atoms (Herbst \& van Dishoeck 2009). Among the discovered
interstellar molecules, COMs are of particular interest for their
potential relation with the origin of the terrestrial life and, more
prosaically, because they are a powerful diagnostic to understand the matter
chemical evolution during the formation of planetary systems like our
own (Caselli \& Ceccarelli 2012).

COMs were first detected in the hot cores associated with high-mass
star forming regions (Cummins et al. 1986; Blake et al. 1986). Since then, they
have been discovered in a variety of environments such as low-mass hot
corinos (Cazaux et al. 2003; Bottinelli et al. 2004; Ceccarelli 2005)
and cold envelopes (Jaber et al. 2014), Galactic Center cold clouds
(Requena-Torrelles et al. 2006), and prestellar cores ({\"O}berg et
al. 2010; Cernicharo et al. 2012; Bacmann et al. 2012).

Currently, models predict that (most of) COMs are synthesised on
the interstellar grain surfaces when they warm up at temperatures
$\geq$30 K, thanks to the enhanced mobility of heavy elements and
radicals on the surface (e.g. Garrod \& Herbst 2006; Garrod et
al. 2009). The detection of COMs in cold ($\leq 20$ K) environments
represents a challenge to those models because the
mentioned mechanism cannot work at those temperatures. Besides, the
process that releases iced species from the grain surfaces into the
gas phase is a problematic issue at such low
temperatures. Recently, Vasyunin \& Herbst (2013) proposed a
mechanism, that they call reactive desorption, in which the
exothermicity of surface chemical reactions make the species to be
desorbed after their formation. Assuming an ejection efficiency of
10\%, this model can approximatively reproduce the abundances observed
in prestellar cores of dimethyl ether (CH$_3$OCH$_3$), acetaldehyde 
(CH$_3$CHO), ketene (H$_2$CCO) and methanol (CH$_3$OH), but it 
fails to reproduce other species such as
formaldehyde (H$_2$CO), methyl formate (HCOOCH$_3$) and
methoxy (CH$_3$O).

Given the key role that the presence of COMs in prestellar cores has
in understanding the general mechanisms of their formation, it is of
paramount importance (i) to have a as complete as possible census of
the COMs present in prestellar cores, (ii) to better characterise where the
COMs emission comes from in these cold objects and, as a consequence,
(iii) to have a better determination of their abundance, at present
obtained by dividing the measured species column density by the total
H$_2$ column density of the core.

In this Letter, we present a complete census of the oxygen bearing
COMs, and other key species useful to constrain COMs formation
processes, in the prestellar core L1544. The census has been obtained
by analysing the unbiased spectral survey in the 3mm band carried out
at the IRAM-30m telescope, within the Large Programme ASAI
(Astrochemical Surveys At Iram)\footnote{http://www.oan.es/asai/}.

\section{The source}

L1544 is a prototypical starless core in the Taurus molecular cloud
complex (d $\sim$ 140 pc) on the verge of the gravitational collapse
(Caselli et al. 2002 and references within). It is characterised by a
central high density (2 $\times$ 10$^6$ cm$^{-3}$), low temperature
($\sim$ 7 K; Crapsi et al. 2007), and high CO depletion, accompanied
by a large degree of molecular deuteration (Caselli et al. 2003;
Crapsi et al. 2005; Vastel et al. 2006).
Its physical and dynamical structure has been recently reconstructed
by Caselli et al. (2012; hereinafter CKB2012) and Keto et al. (2014)
using the numerous existing observations towards L1544.  Among them,
we emphasise the recent detection of water vapour by the Herschel
Space Observatory, the first water detection ever in a prestellar
core, which provided a key information to reconstruct the physical and
chemical structure of L1544 (CKB2012). In fact, the observed water
line shows an inverse P-Cygni profile, characteristic of gravitational
contraction, confirming that L1544 is on the verge of the
collapse. Based on the line shape, CKB2012 predict that
water is largely frozen into the grain mantles in the interior
($\leq$4000 AU) of the L1544 core, where the gaseous H$_2$O abundance
(with respect to H$_2$) is only $\sim10^{-9}$. The low but not zero
level of water vapour is believed to be caused by the photodesorption
of water molecules from the icy mantles by the FUV photons created by
the interaction of cosmic rays with H$_2$ molecules.  Farther away
from the center ($\sim10^4$ AU), where the density is low enough
($\leq 10^5$ cm$^{-3}$) for the photodesorption rate not to be
overcome by the freeze-out rate, the gaseous H$_2$O abundance reaches
$\sim 3\times 10^{-7}$, as predicted by previous models (Dominik
et al. 2005; Hollenbach et al. 2009).

\section{Observations and results}

The observations were performed on July 3rd, 2013 and September 10th,
2013 toward L1544 ($\alpha_{2000} = 05^h04^m17.21^s, \delta_{2000} =
25\degr10\arcmin42.8\arcsec$).  Using the 3 mm Eight Mixer Receivers
(16 GHz of total instantaneous bandwidth per polarisation) and the
fast Fourier Transform Spectrometers with a spectral resolution of 50
kHz (allowing the observation of the inner 1.82 GHz of each band), six
set-ups were needed to cover the full band, namely the frequency range
between 81--110 GHz. Two higher energy level transitions of the C$_3$O
species, at 96.2 and 105.8 GHz respectively, were observed during the
January 2014 run to reach a better rms.

Weather conditions were average with 2 to 3 mm of precipitable water
vapour. System temperatures were between 110 and 180 K, resulting in
an average rms of 4 to 7 mK in a 50 kHz frequency bin. 
in order to obtain a flat baseline, observations were carried out
using a nutating secondary mirror, with a throw of 3 arcmin. No
contamination from the reference position was observed.
Pointing was checked every 1.5 hours on the nearby continuum sources
0439+360 and 0528+134. Pointing errors were always within
3$\arcsec$. 
We adopted the telescope and receiver parameters (main-beam
efficiency, half power beam width, forward effiency) from the values
monitored at IRAM (http://www.iram.fr).
Line intensities are expressed in units of main-beam brightness temperature.

We detected several lines from E- and A- CH$_3$OH and $^{13}$CH$_3$OH,
CH$_3$CHO, t-HCOOH, H$_2$CCO, E-CH$_3$CCH, and C$_3$O. 
All lines, for the first time detected in L1544, are in
emission, except the 107 GHz line of E-CH$_3$OH, which appears in
absorption.  We did not detect any other oxygen bearing COM. Important
non-detections include CH$_3$OCH$_3$, HCOOCH$_3$ and CH$_3$O, which
were, on the contrary, detected in other prestellar cores ({\"O}berg
et al. 2010; Cernicharo et al. 2012; Bacmann et al. 2012).

\begin{figure}
\epsscale{1.0}
%\plotone{L1544_C3O_CH3OH.eps}
\plotone{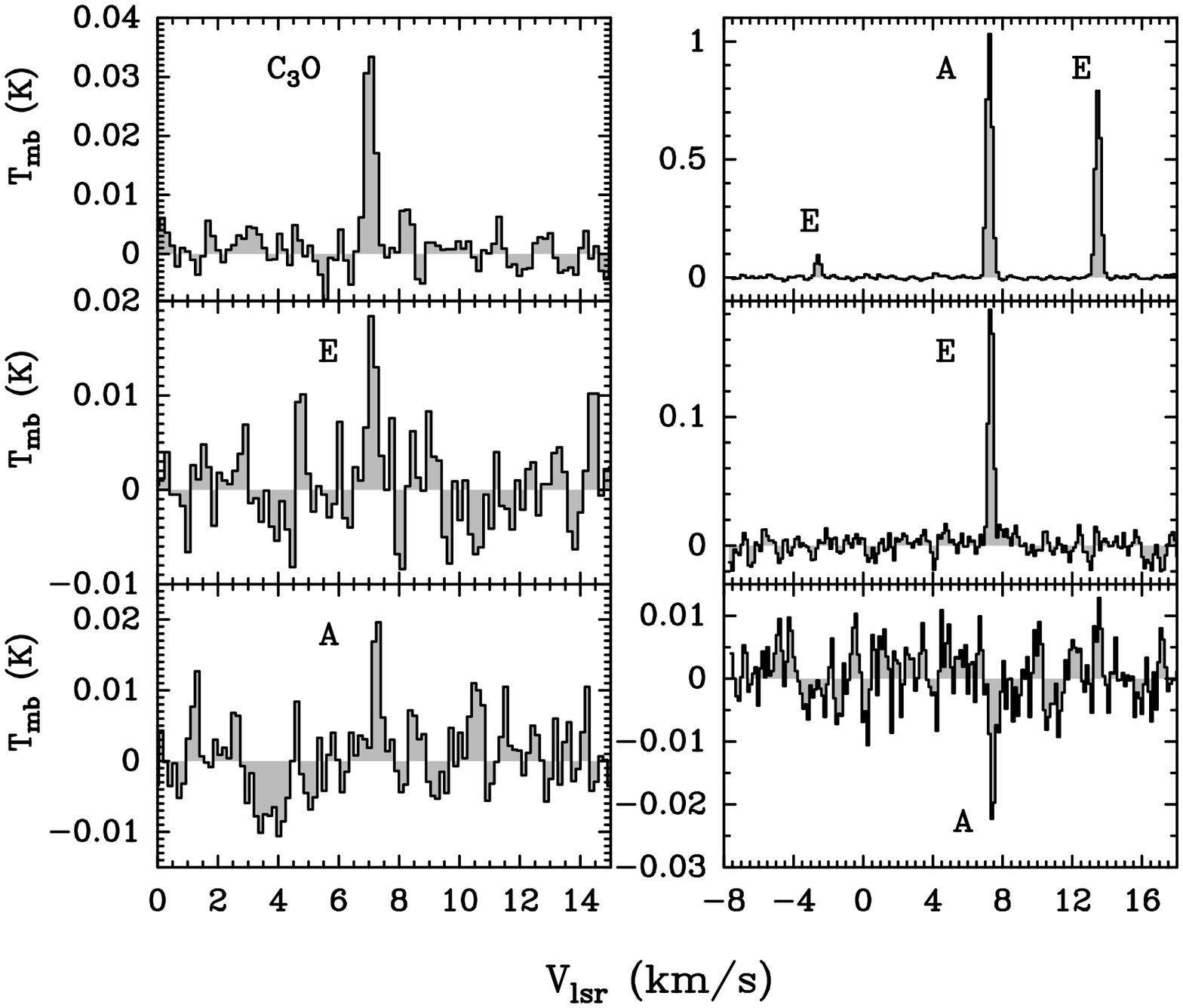}
\caption{Spectra of the detected lines from C$_3$O (upper left pannel)
  and CH$_3$OH (the other panels). The temperatures are main beam
  temperatures. \label{c3o}}
\end{figure}

\begin{figure}
\epsscale{0.8}
%\plotone{tHCOOH_H2CCO_CH3CHO.eps}
\plotone{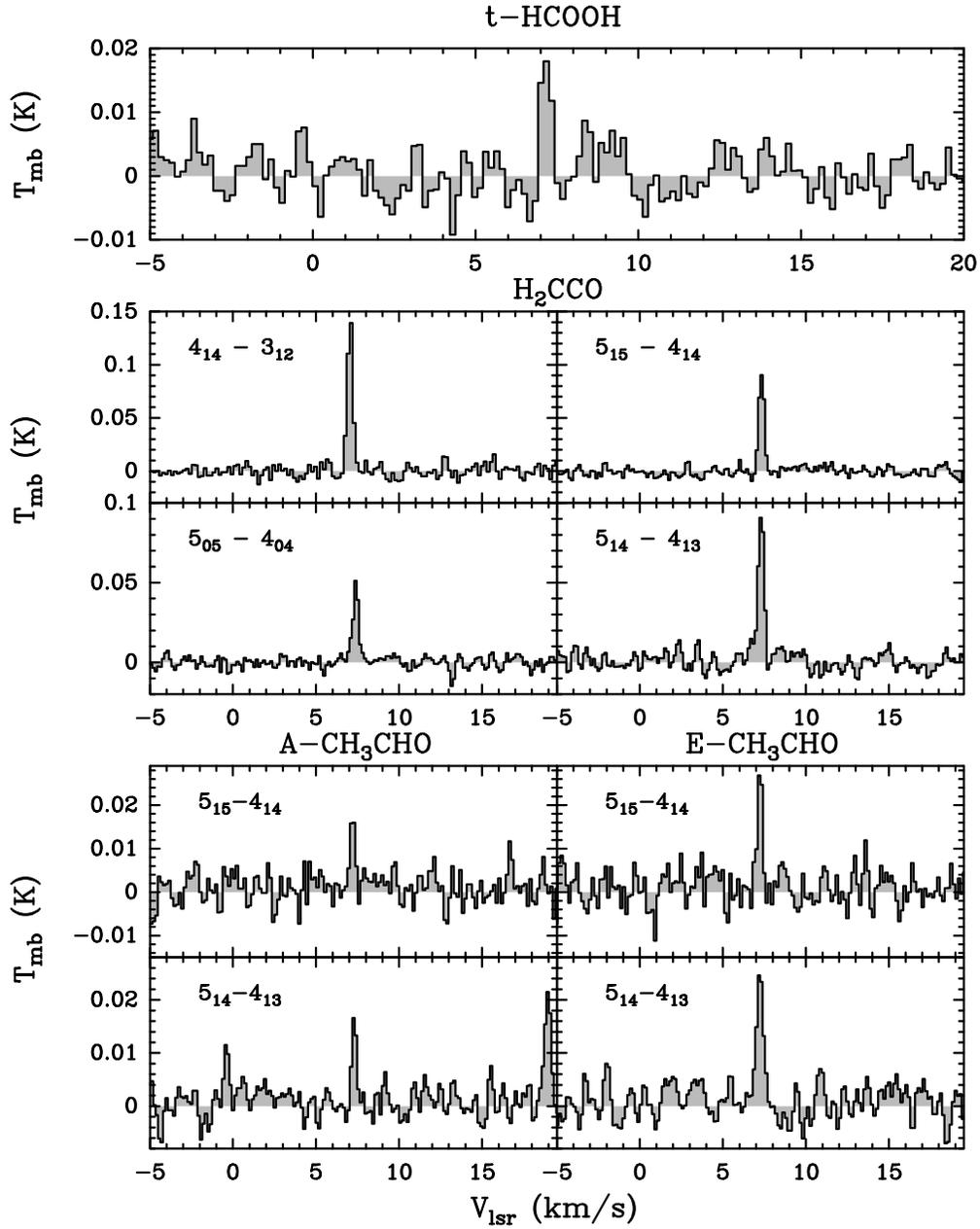}
\caption{Spectra of the detected lines from t-HCOOH (top panel),
  H$_2$CCO (central panel) and CH$_3$CHO (bottom panel). The
  temperatures are main beam temperatures. \label{ch3cch}}
\end{figure}

Figures \ref{c3o} and \ref{ch3cch} show the detected transitions of
C$_3$O, A- and E- CH$_3$OH, CH$_3$CHO, t-HCOOH, H$_2$CCO,
E-CH$_3$CCH. Using the CASSIS\footnote{http://cassis.irap.omp.eu/}
software, the lines have been fitted using the Levenberg-Marquardt
function. The parameters resulting from the fit are listed in Table
\ref{transitions}.

\section{Derivation of the column densities}

We carried out an LTE analysis of the observed line intensities for
all the detected species (except CH$_3$OH, see below).  We assumed
that the source fills the beam\footnote{This is a reasonable
  assumption based on the numerous previous observations towards
  L1544. Nonetheless, since the observed lines have frequencies that
  differ less than 20\%, including a possible filling factor does not
  substantially change the results.} and we carried out a rotational
diagram analysis for the species in which more than one line has been
detected, namely all the detected species except C$_3$O and t-HCOOH.
Specifically, we computed a grid of LTE models for the CH$_3$CHO,
H$_2$CCO and E-CH$_3$CCH lines. Then, for each species, we found the
best fit parameters varying the linewidth, V$_{LSR}$, excitation
temperature (assumed to be the same in all observed lines) and column
density. In the cases of C$_3$O and t-HCOOH, we adopted an excitation
temperature equal to 10 K.  The results of this analysis are reported
in Table \ref{rad_trans}.

\begin{figure}
\epsscale{1.0}
%\plotone{figure-ch3oh.ps}
\plotone{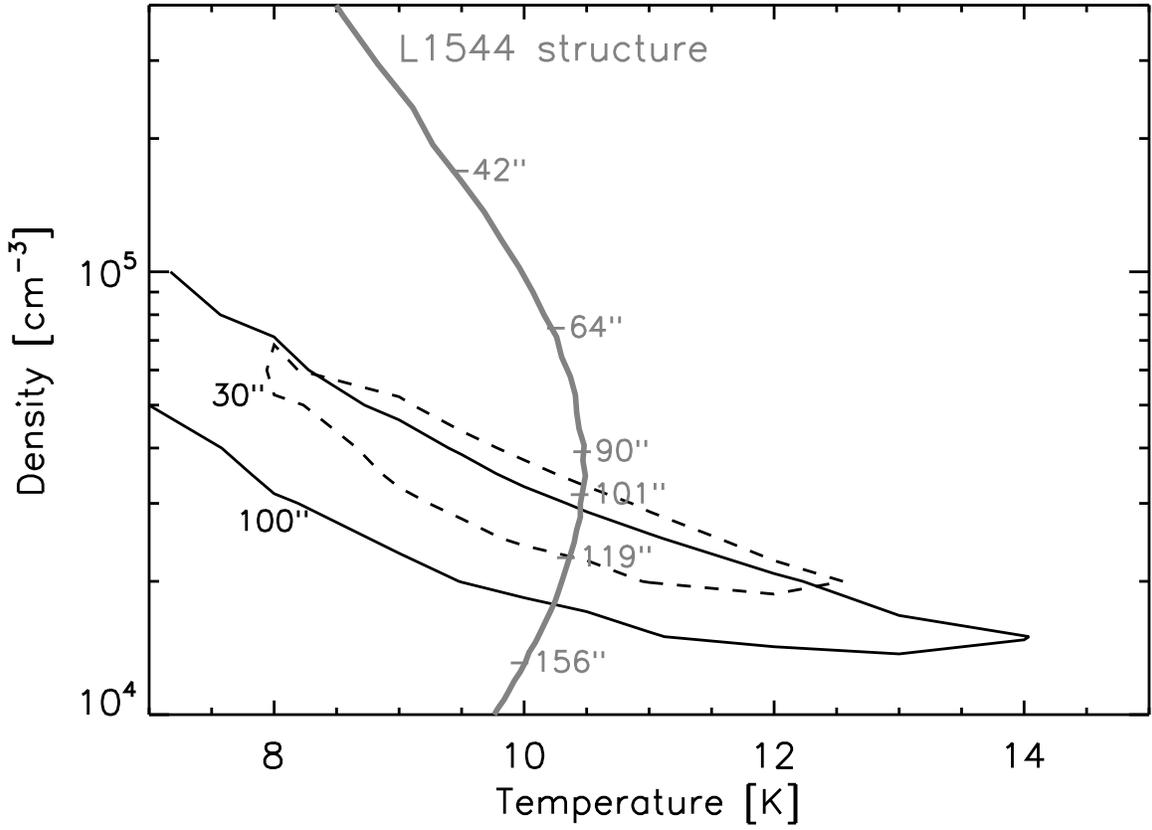}
\caption{Results from the LVG analysis of the CH$_3$OH lines.  The
  black lines show the $\chi^2$=0.8 contour plot for
  N(CH$_3$OH)=$3.8\times10^{13}$ cm$^{-2}$) and a size of $30"$
  (dashed) and N(CH$_3$OH)=$2.6\times10^{13}$ cm$^{-2}$) and size
  $100"$ (solid), the two extremes of the solutions with
  reduced $\chi^2\leq 0.8$. The grey line shows the densities and
  temperatures at different distances from the center (diameter in
  arcsec are marked along the curve) predicted for the structure of
  L1544 by CKB2012.\label{lvg}}
\end{figure}

In the case of CH$_3$OH, we carried out a non-LTE analysis,
considering the A and E forms separately and assuming an abundance
ratio 1:1 between them and the $^{12}$C/$^{13}$C equal to 75. We used
the Large Velocity Gradient (LVG) code described in Ceccarelli et
al. (2003) and the collisional rates by Rabli \& Flower (2010),
retrieved from the BASECOL data base ({\it http://basecol.obspm.fr};
Dubernet et al. 2013). We run a large grid of models varying the
methanol column density from $5 \times 10^{12}$ to $6 \times 10^{13}$
cm$^{-2}$, the density from $1 \times 10^{4}$ to $1 \times 10^{6}$
cm$^{-3}$, the temperature from 7 to 16 K, and the source size from
1$"$ to 200$"$. For each methanol column density, we found the source
size, temperature and density with the minimum reduced $\chi^2$. All
solutions with reduced $\chi^2 \leq 0.8$ are considered good
(equivalent to a probability of 1 $\sigma$).  With this criterium, we
found that CH$_3$OH column densities between 2.6 to 3.8 $\times
10^{13}$ cm$^{-2}$ are consistent with the data. The observed lines
are optically thick for N(CH$_3$OH)$\geq 3.8\times 10^{13}$ cm$^{-2}$
and this allows us to constrain also the emitting size to be larger
than about $30"$. At larger sizes, the beam filling factor becomes
unity and the lines optically thin. The temperature and density are
degenerate, as shown in Fig. \ref{lvg} for the two extreme methanol
column densities. Note, however, that the derived density is $\leq
10^5$ cm$^{-3}$ and the temperature is between 7 and 15 K for a
density between $10^4$ and $10^5$ cm$^{-3}$.  Note also that the best
fit models correctly reproduce the absorption line at 107 GHz.

Finally, the non-detection of lines from CH$_3$OCH$_3$, HCOOCH$_3$ and
CH$_3$O provide upper limits to their column density of $1\times
10^{12}$, $6\times 10^{12}$ and $6\times 10^{11}$, respectively.

\section{Spatial origin and abundances of the COMs}

The analysis of the CH$_3$OH lines provides a strong constraint on the
gas density, which has to be $\leq 10^5$ cm$^{-3}$. This means that the 
bulk of the CH$_3$OH emission does not originate in the
dense interior of the L1544 core, but in a less dense, external
zone. In order to constrain the location of the CH$_3$OH, we overplot
the density and temperature structure derived by CKB2012
on the density-temperature $\chi^2$ contour plot of the LVG analysis.
As can be seen from Fig. \ref{lvg}, the density-temperature $\chi^2$
contours intercept the L1544 structure when the source size is between
100$"$ and 140$"$, namely the CH$_3$OH emitting gas is at a radius
between about 6800 and 9600 AU from the center. At these radii the
CKB2012 structure predicts a density of 
3--1.5 $\times 10^4$ cm$^{-3}$ and a temperature of $\sim$ 10 K. 
The very recent work by Bizzocchi et al. 2014 reach the same conclusion.

The coincidence between the CH$_3$OH emitting region with that where
the gaseous water abundance is the largest (see \S 2) is
remarkable. In other words, the LVG analysis strongly suggests that
methanol has an enhanced abundance where the FUV photons desorb water
molecules. It is, therefore, natural to conclude that methanol has the
same origin, namely it is photodesorbed from the icy
mantles (Andrade et al. 2010).

The identification of the region where the CH$_3$OH lines are emitted
has an obvious impact on the estimate of the methanol abundance, as,
in order to obtain it, one has to divide the measured column density
of Tab. 2 by the H$_2$ column density of the emitting region, namely
$\sim 6$--$4 \times 10^{21}$ cm$^{-2}$, and not the whole L1544 core, which
is $\sim 30$ times larger. Therefore, the (E + A) methanol abundance
is $\sim 6\times 10^{-9}$.

It is natural to assume that the emission from the other detected
COMs comes from the same outer layer of L1544, rather than the core
itself. The rotational temperatures ($\geq 10$ K) in Tab. 2 support
this hypothesis. Therefore, we computed the COMs abundances following
this assumption. The resulting abundances, reported in Tab. 2, vary
from $\sim 5\times 10^{-9}$ for the CH$_3$CCH, the most abundant COM
after methanol, to $\sim 5\times 10^{-11}$ for C$_3$O, the least
abundant.  Finally, the upper limits to the column density of
CH$_3$OCH$_3$, HCOOCH$_3$ and CH$_3$O convert into an upper limit to
their abundance of $2\times 10^{-10}$, $1.5\times 10^{-9}$ and
$1.5\times 10^{-10}$, respectively.

\section{Chemical modeling}

In order to shed light on the formation pathways of the observed COMs,
we run a simple chemical model where we first computed the steady
state abundances of all the species and then studied the effect on the
COMs abundances caused by the injection of grain mantle species into
the gas phase. We do not pretend here to describe the situation with a
self-consistent model, but just to understand what species from the
mantles are needed to reproduce the observed COMs and in what
approximate amount. To make things as simple as possible, we
considered two species: methanol, known to be present in the mantles,
and ethene (C$_2$H$_4$). The latter has not been detected in the ices
(with an abundance larger than $\sim 10^{-7}$), but it is the first
step toward ethane (C$_2$H$_6$), whose oxydation leads to
acetaldehyde, among other oxygen bearing COMs (Charnley 2004). In
addition, we considered the elemental oxygen (in the gas) as a
parameter, in order to account for the amount of oxygen trapped into
the mantle ices.

We used the Nahoon gas-phase chemical model ({\it
  http://kida.obs.u-bordeaux1.fr/models/}: Wakelam et al. 2010), which
computes the chemical evolution of species as a function of time for a
fixed temperature and density. The chemical network kida.uva.2011
({\it http://kida.obs.u-bordeaux1.fr/}) has been updated following
Loison et al. (2014) and includes 6680 reactions over 486 species.  In
our computations we used the following elemental abundances (with
respect to H nuclei):
He=0.07, C=$5\times 10^{-5}$, N=$2\times10^{-5}$ ,
Si=$8\times10^{-9}$, S=$8\times10^{-8}$, Fe=$3\times10^{-9}$,
Na=$2\times10^{-9}$, Mg = $7\times10^{-9}$, Cl=$1\times10^{-9}$,
P=$2\times10^{-10}$. The oxygen elemental abundance was such to have
the following C/O abundance ratios: 0.5, 0.6, 0.7 and 0.8.
For the first step, we adopted a gas and dust temperature of 10 K, an
H density of $2 \times 10^4$ cm$^{-3}$, a visual extinction of 10 mag and
a cosmic rays ionization rate of $3\times 10^{-17}$ s$^{-1}$. We let
the chemical composition evolve until steady state is reached. The
abundances from this first step were then used as initial abundances
for the second step, where we changed the abundance of methanol and
ethene separately. In the following, we discuss the results for the
detected species.

\noindent {\it C$_3$O} is the longest oxygen-bearing carbon chain
observed in the interstellar medium (ISM). So far, very few detections
of the C$_3$O have been reported and only one in a cold dark cloud
core, TMC-1, unusually carbon-rich (Matthews et al. 1984; Kaifu et
al. 2004). Loison et al. (2014) showed that the reactions between
carbon chain molecules and radicals (C$_n$, C$_n$H, C$_n$H$_2$,
C$_{2n+1}$O...) with O atoms produce large enough abundances of the
C$_3$O. We, therefore, started modeling this species as it provides
constraints on the amount of gaseous oxygen. We could reproduce the
observed C$_3$O abundance only when the lowest C/O ratio of 0.5 is
used, namely when a large fraction of oxygen is in the gas phase. This
is in remarkable agreement with the CKB2012 model, which predicts that
a large fraction of water is photodesorbed from the grain mantles,
making the gas oxygen rich.

\noindent {\it CH$_3$OH} is known to be a grain surface product, so
that it is no surprising that a pure gas phase model does not
reproduce the measured abundance. Injecting $\sim1\times10^{-8}$
methanol from the ices provides the observed amount of methanol. This
is a tiny fraction with respect to the methanol formed on the grain
surfaces during the prestellar phase (e.g. Taquet et al. 2012).

\noindent {\it CH$_3$CHO and H$_2$CCO} are overabundant with respect
to the predictions of the step 1 model, by about three and one orders of
magnitude, respectively. However, the injection of ethene with an abundance of
$5\times10^{-9}$ is enough to reproduce the measured abundances through the 
following sequences (Charnley et al. 1992): 
\begin{equation}
%\tiny
%C2H2--H-->C2H4--H-->C2H6--XH+-->C2H7+--e-->C2H5--O-->CH3CHO and C2H4--XH+-->C2H5+--e-->C2H3--O-->H2CCO. 
C_2H_4 \stackrel{H}{\longrightarrow} C_2H_6 \stackrel{XH^+}{\longrightarrow} C_2H_7^+ \stackrel{e}{\longrightarrow} C_2H_5  \stackrel{O}{\longrightarrow} CH_3CHO
\end{equation}
\begin{equation}
%\tiny
C_2H_4 \stackrel{XH^+}{\longrightarrow} C_2H_5^+ \stackrel{e}{\longrightarrow} C_2H_3 \stackrel{O}{\longrightarrow} H_2CCO
\end{equation}

\noindent {\it CH$_3$CCH} is also overabundant with respect to the
predictions of the step 1 model, by about five orders of
magnitudes. The injection of $5\times10^{-9}$ ethene increases the
abundance by three orders of magnitude but it is still not enough to
reproduce the measured abundance.  {\"O}berg et al. (2013) suggested
that an important formation pathway in cold gas is missing for
CH$_3$CCH in the models, as reinforced by our new measurements.

\noindent {\it t-HCOOH} is the only species underabundant with respect
to the predicted value of the step 1 model, by about one order of
magnitude. This suggests that, in this case, routes of destruction of
this species are probably missing.

Finally, the non detection of CH$_3$OCH$_3$, HCOOCH$_3$ and
  CH$_3$O is consistent with our and Vasyunin \& Herbst (2013)
  model predictions.

\section{Discussion and conclusions}

The three major results of this work are:
\begin{enumerate}
\item The detection of CH$_3$OH, CH$_3$CHO, t-HCOOH, H$_2$CCO, CH$_3$CCH,
and C$_3$O. Their abundances are equal to $\sim 1\times10^{-11}$
(CH$_3$CHO and t-HCOOH), $\sim 5\times10^{-11}$ (C$_3$O), $\sim
1\times10^{-12}$ (H$_2$CCO), $\sim 5\times10^{-9}$ (CH$_3$CCH), and
$\sim 6\times10^{-9}$ (CH$_3$OH).  No other oxygen bearing COM has
been detected, including CH$_3$OCH$_3$, HCOOCH$_3$ and CH$_3$O,
species that have been detected in other prestellar cores (\S 1). The
upper limit to their abundances is $\leq 10^{-9}-10^{-10}$.
\item The gas has to be oxygen rich (C/O$\sim$0.5) and an injection of a
relatively small amount of methanol ($\sim 10^{-8}$) and ethene ($\sim
5 \times10^{-9}$) from the ices is enough to reproduce the observed
abundances of CH$_3$OH, CH$_3$CHO, H$_2$CCO, and C$_3$O. On the
contrary, the model overestimates the abundance of t-HCOOH and
underestimates that of CH$_3$CCH.
\item Probably, the most important result is the discovery that the
CH$_3$OH emission comes from the border of the L1544 core, at $\sim
8000$ AU, in a region where the ices are desorbed through non-thermal processes. 
We suggest
that very likely also the other detected COMs have the same
origin. The large rotational temperatures support this hypothesis.
\end{enumerate}

L1544 is the third prestellar core, after B1-b and L1689B, where COMs
have been detected ({\"O}berg et al. 2010; Cernicharo et al. 2012;
Bacmann et al. 2012). In the three cores, acetaldehyde, ketene and
acid formic have similar column densities (within a factor 3--4). The
same applies to methanol in L1544 and B1-b. Finally, also the column
densities of methyl formate and dimethyl ether are similar in B1-b and
L1689B, and consistent with our upper limits.  This may lead to
suppose that also in B1-b and L1689B the observed COM line emission
originates in an outer layer where desorption is not fully
compensate by the freeze-out rather than in the core. If this is the
case, the COMs chemistry in prestellar cores may be driven by the
non-thermal desorption of simple ice components, namely hydrogenated species
like methanol and ethene, and not necessarily by other processes
(e.g. Cernicharo et al. 2012; Vasyunin \& Herbst 2013).

In conclusion, L1544 is a core on the verge of collapsing (Caselli
et al. 2002; Keto et al. 2014). The center of the core is very cold
and dense, and all species are largely frozen onto the grain mantles
(CKB2012). There is, however, a region, at $\sim$8000 AU from the
center, where FUV photons (the product of the interaction of the
cosmic rays with the H$_2$ molecules according to CKB2012) photodesorb
a tiny fraction of the frozen water and possibly methanol and ethene. These
molecules react with other species in the gas phase and produce a
detectable amount of acetaldehyde, ketene and tricarbon monoxide. It
is possible, but not demonstrated, that a similar situation occurs
also in other prestellar cores, for example in B1-b and L1689B, where
previous observations have detected COMs with similar column
densities. If this is the case, the quoted abundances may need some
revision. Also in these cases, the COMs chemistry may be driven by the 
desorption of simple hydrogenated species from the ices. Higher
spatial resolution observations are necessary to settle the issue and
better constrain the formation routes of COMs in prestellar cores.

\begin{acknowledgements}
We thank P. Caselli, J.-C. Loison and M. Ruaud for helpful discussion. 
\end{acknowledgements}

{\it Facilities:} \facility{IRAM}, \facility{CASSIS}.

\clearpage

\begin{deluxetable}{cccccccccc}
\tabletypesize{\scriptsize}
%\rotate
\tablewidth{0pt}
\tablecaption{Spectroscopic parameters of C$_3$O  and the COMs detected toward L1544 and results from the line Gaussian fits. 
The rms on a 50 kHz frequency bin was computed over a range of 40 km/s. The error in the main-beam temperature T$_{\rm mb}$ does not include the calibration.
Note that the A and E forms for CH$_3$OH are taken from the CASSIS database. 
The line labelled $^a$ is in absorption.\label{transitions}}
\tablehead{
\colhead{Species} & \colhead{Database} & \colhead{QN} & \colhead{Frequency}  & \colhead{E$_{\rm up}$}    & \colhead{A$_{\rm ij}$}   & \colhead{rms}  & \colhead{T$_{\rm mb}$}    & \colhead{FWHM}             & \colhead{V$_{LSR}$} \\  
               &   &        &   \colhead{(GHz)}      &    \colhead{(K)}                  & \colhead{(s$ ^{-1}$)}     &  \colhead{(mK)}         &   \colhead{(mK)}         &  \colhead{(km s$^{-1}$)}         &  \colhead{(km s$^{-1}$)}
}
\startdata
C$_3$O  & 52501 &   9 -- 8      & 86.59368      & 20.78 &  2.05 $\times$ 10$^{-5}$ &  3.5  & 36.8 $\pm$ 3.5  & 0.39 $\pm$ 0.04 & 7.04 $\pm$ 0.02\\
               &            &  10 -- 9      & 96.21462     & 25.40 &  2.82 $\times$ 10$^{-5}$ &  6.1  & \nodata  &\nodata & \nodata\\
               &            &  11 -- 10    & 105.83536 & 30.48 &  3.77 $\times$ 10$^{-5}$ &  4.9  & \nodata   & \nodata & \nodata\\    
 A-CH$_3$CHO  & 44003  & 515-414  & 93.58091  & 15.75 & 2.53 $\times$ 10$^{-5}$ & 3.5   & 17.8 $\pm$ 3.4  & 0.33 $\pm$ 0.07 & 7.20 $\pm$ 0.03\\ 
                           &              & 505-404  & 95.96346  & 13.84 & 2.84 $\times$ 10$^{-5}$ & 5.9   & 25.6 $\pm$ 3.6  & 0.81 $\pm$ 0.13 & 7.35 $\pm$ 0.06\\                    
                           &              & 514-413  & 98.90094  & 16.51 & 2.99 $\times$ 10$^{-5}$ & 2.7   & 17.2 $\pm$ 2.6  & 0.31 $\pm$ 0.05 & 7.30 $\pm$ 0.02\\                                                       
 E-CH$_3$CHO  & 44003  & 515-414  & 93.59523  & 15.82 & 2.53 $\times$ 10$^{-5}$ & 3.9   & 29.3 $\pm$ 3.8  & 0.33 $\pm$ 0.05 & 7.23 $\pm$ 0.02\\
                           &              & 505-404  & 95.94744  & 13.93 & 2.84 $\times$ 10$^{-5}$ & 5.8   & 50.5 $\pm$ 8.4  & 0.24 $\pm$ 0.05 & 7.20 $\pm$ 0.01\\
                           &              & 514-413  & 98.86331  & 16.59 & 2.99 $\times$ 10$^{-5}$ & 3.2   & 24.7 $\pm$ 2.4  & 0.49 $\pm$ 0.05 & 7.23 $\pm$ 0.02\\              
t-HCOOH           & 46506   & 414-313  & 86.54619  & 13.57  & 6.35 $\times$ 10$^{-6}$ & 3.4   & 19.9 $\pm$ 3.3  & 0.34 $\pm$ 0.07 & 7.17 $\pm$ 0.03\\ 
H$_2$CCO  & 42501   & 413-312  & 81.58623   & 22.84  & 5.33 $\times$ 10$^{-6}$ & 5.8    & 147.2 $\pm$ 5.6  & 0.38 $\pm$ 0.02 & 7.05 $\pm$ 0.01\\                                   
                    &              & 515-414  & 100.09451  & 27.46  & 1.03 $\times$ 10$^{-5}$ & 4.2   & 95.5 $\pm$ 3.7  & 0.39 $\pm$ 0.02 & 7.32 $\pm$ 0.01\\                                   
                    &              & 505-404  & 101.03663  & 14.55  & 1.10 $\times$ 10$^{-5}$ & 3.5     & 49.1 $\pm$ 3.0  & 0.42 $\pm$ 0.03 & 7.39 $\pm$ 0.01\\                                   
                    &              & 514-413  & 101.98143  & 27.74  & 1.09 $\times$ 10$^{-5}$ & 5.3     & 95.3 $\pm$ 4.5  & 0.41 $\pm$ 0.02 & 7.30 $\pm$ 0.01\\                                   
E-CH$_3$CCH & 40592  & 52-42  &  85.45077  & 41.21 & 1.70 $\times$ 10$^{-6}$ & 3.8     & 70.8 $\pm$ 3.4  & 0.46 $\pm$ 0.03 & 7.18 $\pm$ 0.01\\
                         &             & 51-41  &  85.45567  & 19.53 & 1.95 $\times$ 10$^{-6}$ & 3.6     & 740.8 $\pm$ 15.9  & 0.43 $\pm$ 0.01 & 7.18 $\pm$ 0.01\\
                         &             & 50-40  &  85.45730  & 12.30 & 2.03 $\times$ 10$^{-6}$ & 3.6     & 740.2 $\pm$ 10.3  & 0.46 $\pm$ 0.01 & 7.18 $\pm$ 0.01\\
                         &             & 62-52  &  102.54014  & 46.13 & 3.16 $\times$ 10$^{-6}$ & 4.4     & 61.1 $\pm$ 3.5  & 0.46 $\pm$ 0.03 & 7.40 $\pm$ 0.01\\
                         &             & 61-51  &  102.54602  & 24.45 & 3.46 $\times$ 10$^{-6}$ & 4.4     & 612.6 $\pm$ 10.6  & 0.47 $\pm$ 0.01 & 7.38 $\pm$ 0.01\\
                         &             & 60-50  &  102.54798  & 17.23 & 3.56 $\times$ 10$^{-6}$ & 4.5     & 632.4 $\pm$ 10.8  & 0.47 $\pm$ 0.01 & 7.38 $\pm$ 0.01\\
E-CH$_3$OH   & 32083  & 5 -1 0 - 4 0 0  &  84.52117  & 32.49 & 1.97 $\times$ 10$^{-6}$ & 4.1  & 18.9 $\pm$ 4.0  & 0.37 $\pm$ 0.09 & 7.12 $\pm$ 0.04\\
                         &             & 2 -1 0 - 1 -1 0  &  96.73936  & 4.64 & 2.55 $\times$ 10$^{-6}$ & 6.8  & 805.0 $\pm$ 6.0  & 0.38 $\pm$ 0.01 & 7.24 $\pm$ 0.01\\
                         &             & 2 0 0 - 1 0 0  &  96.74454  & 12.19 & 3.40 $\times$ 10$^{-6}$ & 6.1  & 95.8 $\pm$ 5.7  & 0.34 $\pm$ 0.02 & 7.25 $\pm$ 0.01\\
                         &             & 0 0 0 - 1 -1 0  &  108.89394  & 5.23 & 1.47 $\times$ 10$^{-5}$ & 8.1  & 195.2 $\pm$ 7.1  & 0.34 $\pm$ 0.01 & 7.33 $\pm$ 0.01\\
E-$^{13}$CH$_3$OH  & 32502 & 2 -1 0 - 1 -1 0  &  94.40516  & 4.53 & 2.38 $\times$ 10$^{-6}$ & 6.8  & 20.1 $\pm$ 6.0  & 0.45 $\pm$ 0.15 & 7.02 $\pm$ 0.06\\  
A-CH$_3$OH   & 32093 & 2 0 + 0 - 1 0 + 0  &  96.74137   & 6.97   & 3.40 $\times$ 10$^{-6}$ & 6.6  & 1047.5 $\pm$ 6.1  & 0.39 $\pm$ 0.01 & 7.24 $\pm$ 0.01\\
                        &             & 2 1 - 0 - 1 1 - 0   &  97.58280    & 21.56 & 2.62 $\times$ 10$^{-6}$ & 5.6  & 21.8 $\pm$ 5.7  & 0.29 $\pm$ 0.09 & 7.27 $\pm$ 0.04\\    
                        &             & 3 1 + 0 - 4 0 + 0  &  107.01383  & 28.35 & 3.06 $\times$ 10$^{-6}$ & 4.4  & 22.3 $\pm$ 3.5$^a$  & 0.41 $\pm$ 0.08 & 7.45 $\pm$ 0.03\\          
A-$^{13}$CH$_3$OH  & 32502  & 2 0 + 0 - 1 0 + 0  &  94.40713   & 6.80   & 3.17 $\times$ 10$^{-6}$ & 6.6  & 25.9 $\pm$ 6.1  & 0.40 $\pm$ 0.15 & 7.23 $\pm$ 0.05\\
\enddata
\end{deluxetable}

\begin{deluxetable}{lccccc}
\tablewidth{0pt}
\tablecaption{Column densities and abundances of COMs in L1544. 
Notes: a) LTE minimisation procedure; b) assumed T$_{ex}$=10 K; c) non-LTE LVG analysis (see text). 
The abundances $x$ have been computed assuming N(H$_2$)=$5\times 10^{21}$ cm$^{-2}$.
The final error on the column densities and abundances is about a factor 2.
\label{rad_trans}}
\tablehead{
\colhead{Species}  & \colhead{T$_{ex}$}   & \colhead{T$_k$}      &  \colhead{n$_{H_2}$}                    & \colhead{N}                  &  \colhead{$x$}          \\
                             & \colhead{(K)}          & \colhead{(K)}            & \colhead{(cm$^{-3}$)}  & \colhead{(cm$^{-2}$)}  &      
}
\startdata
C$_3$O$^b$           &  10                 &       &                          & $2 \times 10^{11}$  & $5\times10^{-11}$\\
CH$_3$CHO$^a$     &  17 $\pm$ 1   &       &                          & $5 \times 10^{11}$  & $1\times10^{-10}$\\
t-HCOOH$^b$        &  10                 &       &                          & $5 \times 10^{11}$  & $1\times10^{-10}$\\
H$_2$CCO$^a$       &  27  $\pm$ 1  &       &                          & $5 \times 10^{12}$  & $1\times10^{-9}$\\
E-CH$_3$CCH$^a$  &  11 $\pm$ 2   &       &                          & $2 \times 10^{13}$  & $5\times10^{-9}$\\
CH$_3$OH$^c$       &                      &  10 &  $3\times 10^4$  & $3\times 10^{13}$ & $6\times10^{-9}$\\ \hline
%E-CH$_3$OH$^c$    &                      &  10 &  $3\times 10^4$  & $1.5\times 10^{13}$ & $4\times10^{-9}$\\
%A-CH$_3$OH$^c$   &                       &  10 &  $3\times 10^4$  & $1.5\times 10^{13}$ & $4\times10^{-9}$\\ \hline
CH$_3$OCH$_3$$^b$ & 10               &      &                           & $\leq 1\times 10^{12}$ & $\leq 2\times 10^{-10}$ \\
HCOOCH$_3$$^b$     & 10               &      &                           & $\leq 6\times 10^{12}$ & $\leq 1.5\times 10^{-9}$ \\
CH$_3$O$^b$           & 10                &      &                           & $\leq 6\times 10^{11}$ & $\leq 1.5\times 10^{-10}$ \\
\enddata
\end{deluxetable}

\end{document}